\documentclass[sigconf,authorversion]{acmart}
\setcopyright{acmcopyright}

\usepackage{acmart-taps}

\usepackage[inline]{enumitem}
\usepackage{cleveref}
\newcommand{\etal}{\textit{et al.}}
\newcommand{\ourmethod}{\textit{EditIQ}}

\newlist{inlinelist}{enumerate*}{1}
\setlist[inlinelist]{label=(\roman*)}

\AtBeginDocument{%
  }

\copyrightyear{2025}
\acmYear{2025}
\setcopyright{acmlicensed}
\acmConference[IUI '25]{30th International Conference on Intelligent User Interfaces}{March 24--27, 2025}{Cagliari, Italy}
\acmBooktitle{30th International Conference on Intelligent User Interfaces (IUI '25), March 24--27, 2025, Cagliari, Italy}
\acmDOI{10.1145/3708359.3712113}
\acmISBN{979-8-4007-1306-4/25/03}

\begin{document}


\title[EditIQ: Automated Cinematic Editing]{EditIQ: Automated Cinematic Editing of Static Wide-Angle Videos via Dialogue Interpretation and Saliency Cues}



\author{Rohit Girmaji}
\authornotemark[1]
\email{rohit.girmaji@research.iiit.ac.in}
\affiliation{%
  \institution{CVIT, IIIT Hyderabad, India}
  \country{}
}

\author{Bhav Beri}
\authornotemark[1]
\email{bhav.beri@research.iiit.ac.in}
\affiliation{%
  \institution{CVIT, IIIT Hyderabad, India}
 \country{}
}

\author{Ramanathan Subramanian}
\email{ramanathan.subramanian@ieee.org}
\affiliation{
    \institution{University of Canberra, Australia}
 \country{}
}

\author{Vineet Gandhi}
\email{vgandhi@iiit.ac.in}
\affiliation{%
  \institution{CVIT, IIIT Hyderabad, India}
\country{}
}


\begin{abstract}

We present \ourmethod, a completely automated framework for cinematically editing scenes captured via a stationary, large field-of-view and high-resolution camera. From the static camera feed, \ourmethod~ initially generates multiple virtual feeds, emulating a team of cameramen. These virtual camera shots termed \textit{rushes} are subsequently assembled using an automated editing algorithm, whose objective is to present the viewer with the most vivid scene content. To understand key scene elements and guide the editing process, we employ a two-pronged approach: (1) a large language model (LLM)-based dialogue understanding module to analyze conversational flow, coupled with (2) visual saliency prediction to identify meaningful scene elements and camera shots therefrom. We then formulate cinematic video editing as an energy minimization problem over shot selection, where cinematic constraints determine shot choices, transitions, and continuity. \ourmethod~ synthesizes an aesthetically and visually compelling representation of the original narrative while maintaining cinematic coherence and a smooth viewing experience.
Efficacy of \ourmethod~ against competing baselines is demonstrated via a psychophysical study involving twenty participants on the \textit{BBC Old School} dataset plus eleven theatre performance videos.
Video samples from \ourmethod~ can be found at \url{https://editiq-ave.github.io/}.
\end{abstract}


\begin{teaserfigure}
  \includegraphics[width=\textwidth]{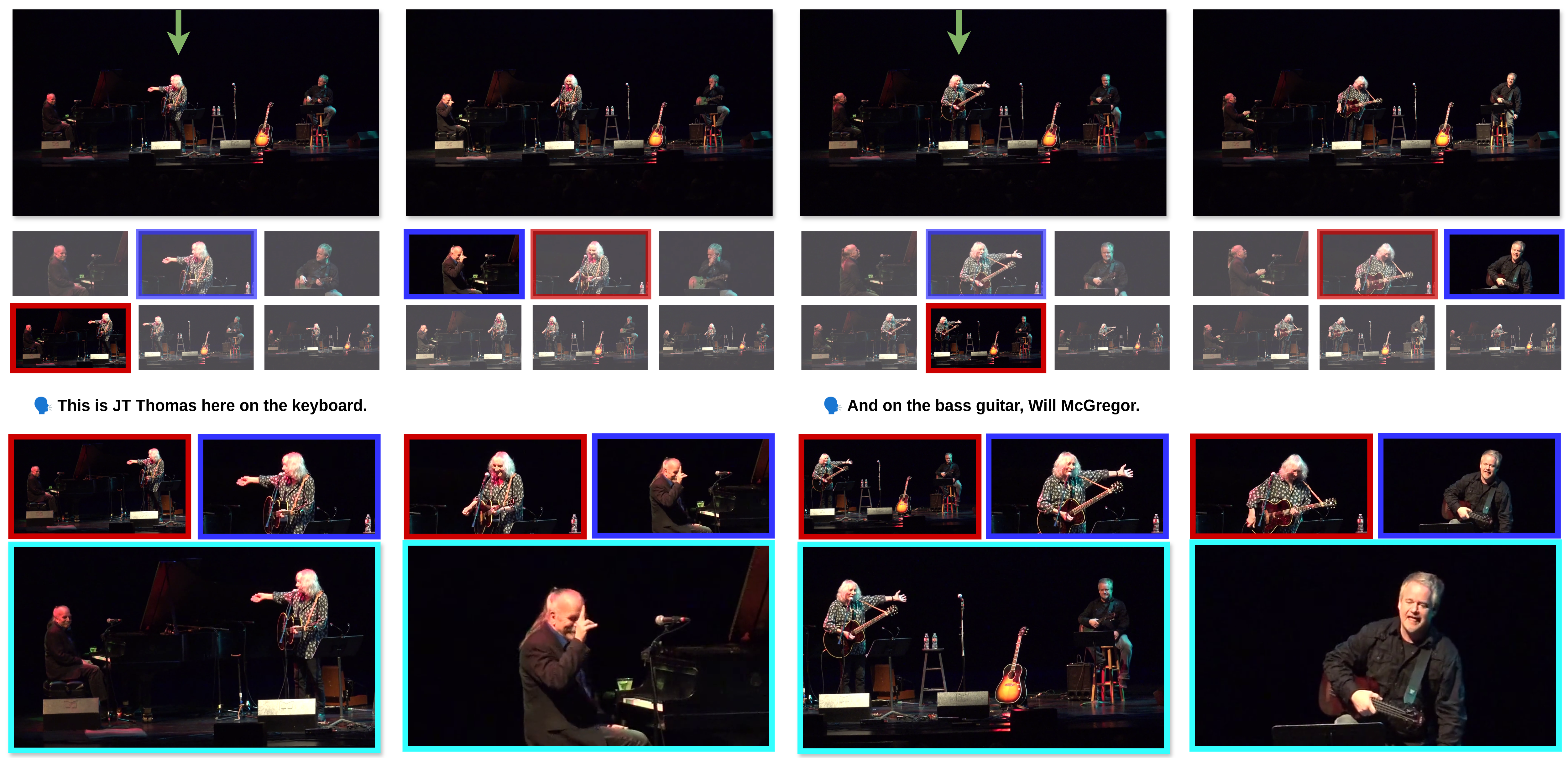}
  \caption{We present \ourmethod, an automated video editing pipeline based on dialogue understanding using LLMs and visual understanding via video saliency. First row presents original video frames input to the pipeline, which generates multiple \textit{rushes} (depicted in the next two rows). The speaker is denoted by a green arrow in the original frame and the transcript below 3rd row. LLMs are employed to analyze the scene’s narrative, guiding shot selections (highlighted in red on the left for each frame in 4th row). Simultaneously, saliency analysis captures prominent visual scene content, giving alternate shot selections (shown in blue on 4th row right for each frame). Combining the language and visual-based scene understanding results generates optical video shots captured in the 5th row.}
  \label{fig:teaser}
\end{teaserfigure}


\maketitle

\section{INTRODUCTION}


Professional video production of an event like a theatre performance or a quiz show usually requires a team of experienced camera operators to film the scene from multiple angles. These multi-camera recordings, termed \textit{rushes}, are later compiled through careful manual editing to craft a coherent narrative designed to enhance audience engagement and viewing experience. The editing of these performances is typically performed in chronological order, with the process primarily focused on selecting the most appropriate rush at each moment. However, filming in confined spaces, such as live theatre performances, presents unique challenges, including limited vantage points, the impossibility of performing retakes, and the impracticality of maneuvering bulky equipment, making the task both difficult and demanding. Consequently, the need for (i) a skilled camera crew, (ii) multiple cameras and supporting equipment, and (iii) expert editors significantly increases the complexity and cost of the video production process.

Consequently, production houses use a wide-field-of-view static camera positioned at a distance suitable to capture the entire stage. This method is common due to its ease of implementation and ability to capture the entire scene. While effective for archival purposes, wide-angle visuals are ineffective in engaging the audience. The distant camera feed conveys the entire scene but fails to capture close-up details like facial expressions and emotions, which are central to cinematic storytelling. As renowned film editor Thelma Schoonmaker once said, ``Close-ups reveal the soul of the character, engaging the audience in a profound way.''


Prior automated video editing efforts have sought to transform static wide-angle footage into more engaging content using machine learning and optimization techniques. The goal is to reduce costs and complexity of production, while still delivering quality content. To reduce the reliance on multiple cameramen, Gandhi~\etal~\cite{vcs,gandhi2015computational} proposed a framework for automatically generating multiple clips suitable for video editing by simulating pan-tilt-zoom camera movements within the frame of a single static camera. Moorthy~\etal~\cite{gazed} demonstrated that efficient camera selection can be achieved by leveraging eye-gaze data from users. Their work assumes that humans inherently focus on salient scene aspects, and that gaze can serve as a proxy to localize key scene elements. Though their method produces impressive results, its usefulness is restricted due to reliance on gaze data, which may not always be available. 

To eliminate the need for auxiliary user data to perform video editing, we present \ourmethod, a fully automated multi-camera video production pipeline for staged events, using footage from one or more wide-angle, stationary static cameras. We leverage prior efforts~\cite{vcs} for simulating multiple virtual cameras, and our focus is on automating the camera selection process. Optimal camera selection demands a nuanced understanding of the scene, capturing key elements such as dialogue, the speaker, and the actors' actions and reactions. Reaction shots, in particular, are vital in editing as they convey the emotional tone of the scene, enhancing the viewer’s engagement with the scene and their perceptual understanding. While the human gaze may naturally track key scene elements, automating this intricate process poses a significant challenge.

\ourmethod~ primarily seeks to leverage advancements in large language models (LLMs) for automated video editing. Recent studies have highlighted strong LLM capabilities for understanding key scene elements such as emotions~\cite{vishnubhotla2024emotion}, entailment~\cite{zhou2024pop}, and co-reference resolution~\cite{manikantan2024major}. Our work is the first to demonstrate that these models can be effectively utilized to guide camera selection based on narratives and conversations, and determining which scene elements should be visually emphasized. \textit{E.g.}, in the exemplar scene illustrated in \Cref{fig:teaser}, a camera following the speaker (or an audio source) would focus solely on the lead singer, missing the non-verbal reactions of other musicians as they are introduced. An LLM instead recognizes that when the lead singer says ``J T Thomas on the keyboard,'' the visual attentional focus should shift to the addressed person. 

LLMs nevertheless struggle to capture scene actions not explicitly referenced in dialogue, and multimodal LLMs typically perform frame-level processing, making them less effective at understanding temporal actions. To overcome this limitation, we additionally utilize a saliency prediction architecture trained to model human gaze, identifying key areas of importance in a scene. Specifically, we extend a spatio-temporal action localization backbone~\cite{pan2021actor} for video saliency prediction~\cite{dhf1k}. Again referring to \Cref{fig:teaser}, the visual saliency network accurately captures key actions, like the keyboard player's salute or the lead singer's hand movement, while referring to the bass guitarist.

Once the LLM and visual saliency pipelines enable the identification of key scene elements (and corresponding video shots), we formulate camera shot selection as a discrete optimization problem, where one among the rushes is selected for viewer presentation at each time-frame. Speaker information (obtained from off-the-shelf detection models), LLM predictions, and saliency outputs serve as three unary potential terms in the cost matrix. Akin to~\cite{gazed}, these potentials are combined with additional constraints that model cinematic editing principles, such as avoiding jump cuts (causing jarring transitions), maintaining rhythm (consistent pacing of transitions), avoiding transient shots, and ensuring proper framing (to prevent cutting off actors). This optimization is then solved using dynamic programming.

To validate \ourmethod, we performed a psychophysical study with 20 participants, comparing multiple edited versions of performance recordings from the BBC Old School Dataset (BBC-OSD) \cite{bbc_osd}, which captures a quiz show plus 11 theatre sequences. Our editing strategy surpasses several competing approaches, including random editing, wide-shot framing, and speaker detection-based editing. On BBC-OSD, \ourmethod~achieves an output that is considerably close to the expert human edit. Our research contributions include:

{\bf (1) Semantic shot potentials:} We use novel LLM and visual saliency-based predictions as potentials to quantify the importance of multiple rushes generated from the original recording. LLMs provide dialogue-based visual cues, while saliency augments information regarding scene actions. 

{\bf (2) Fully automated editing pipeline:} The potentials derived from active speaker detection, LLM predictions, and visual saliency outputs are combined with cinematic constraints, framing camera shot selection as a discrete optimization problem. This process is fully automated, requiring only the wide-angle video and scene information to synthesize the edit. \ourmethod~can edit a 2-minute video featuring five performers in just 2 minutes on a PC equipped with Nvidia 4090Ti GPU. In contrast, manual editing is significantly more onerous and time-consuming.

{\bf (3) Comprehensive Evaluation:} We evaluate our method on the professionally curated BBC-OSD dataset \cite{bbc_osd}, which is specifically designed to assess the automated editing of wide-angle recordings. We compiled an additional set of 11 high-quality 4K theatre sequences to add variety to our evaluation. Results from a comprehensive user study indicate that \ourmethod~outputs are preferred by users in terms of narrative effectiveness, preservation of emotions and actions, and overall viewing experience.

\section{RELATED WORK}

\subsection{Editing through automated crops}

Working with high-resolution footage like 4K or 8K opens up a variety of creative possibilities for editing, especially for tracking and zooming in on specific video segments. Dynamic cropping has demonstrated notable success in the domains of video retargeting and video stabilization. Automated video retargeting focuses on adjusting content to a specified aspect ratio by dynamically selecting cropping windows. Previous attempts to address this issue have drawn on user annotations~\cite{10.1145}, motion or saliency information~\cite{wang2010motion,liu2006video}, and gaze tracking~\cite{rachavarapu2018watch}. Grundmann~\etal~\cite{5995525} proposed an algorithm for automatically applying constrainable, L1-optimal camera paths to generate stabilized videos by removing undesired motions. 

In automated production systems based on cropping, early efforts primarily focused on lectures and presentations~\cite{heck2007virtual,zhang2008automated}. These approaches typically rely on rule-based editing processes confined to controlled settings, featuring a single presenter in front of a chalkboard or slide screen. In the context of sports, Schäfer ~\etal~\cite{schafer2010ultra} introduced a system that enables the visualization of user-specific cropping windows within an ultra-high-resolution feed. Carr~\etal~\cite{carr2013hybrid} used virtual pan-tilt-zoom (PTZ) cropping over a robotic player following a camera for editing of Basketball games. Additionally, virtual PTZ movements have been explored for cinematographic editing in panoramic and 360-degree videos~\cite{su2016pano2vid,tang2019joint,10.1145/2744411}.

In contrast to traditional pan-and-scan-like content editing, our work focuses on simulating a multicamera workflow with automated camera selection. Gandhi~\etal~\cite{vcs} proposed a method for simulating virtual PTZ cameras by shifting a cropping window within wide-angle recordings, aiming to replicate cameraman-like movements using L1-norm-based optimization~\cite{5995525}. Building on this approach, our work introduces automatic selection among these simulated virtual cameras. 

The aforementioned methods also relate and benefit from advancements in computer vision and machine learning techniques, such as object detection~\cite{yolo_repo}, action recognition and localization~\cite{acarnet,slowfast}, person tracking~\cite{botsort,cao2023observation}, pose estimation~\cite{yolov8}, video saliency prediction~\cite{dhf1k,vinet}, and head pose estimation~\cite{kao2023toward}. 

\begin{figure*}[t]
\centering
\includegraphics[width=\textwidth]{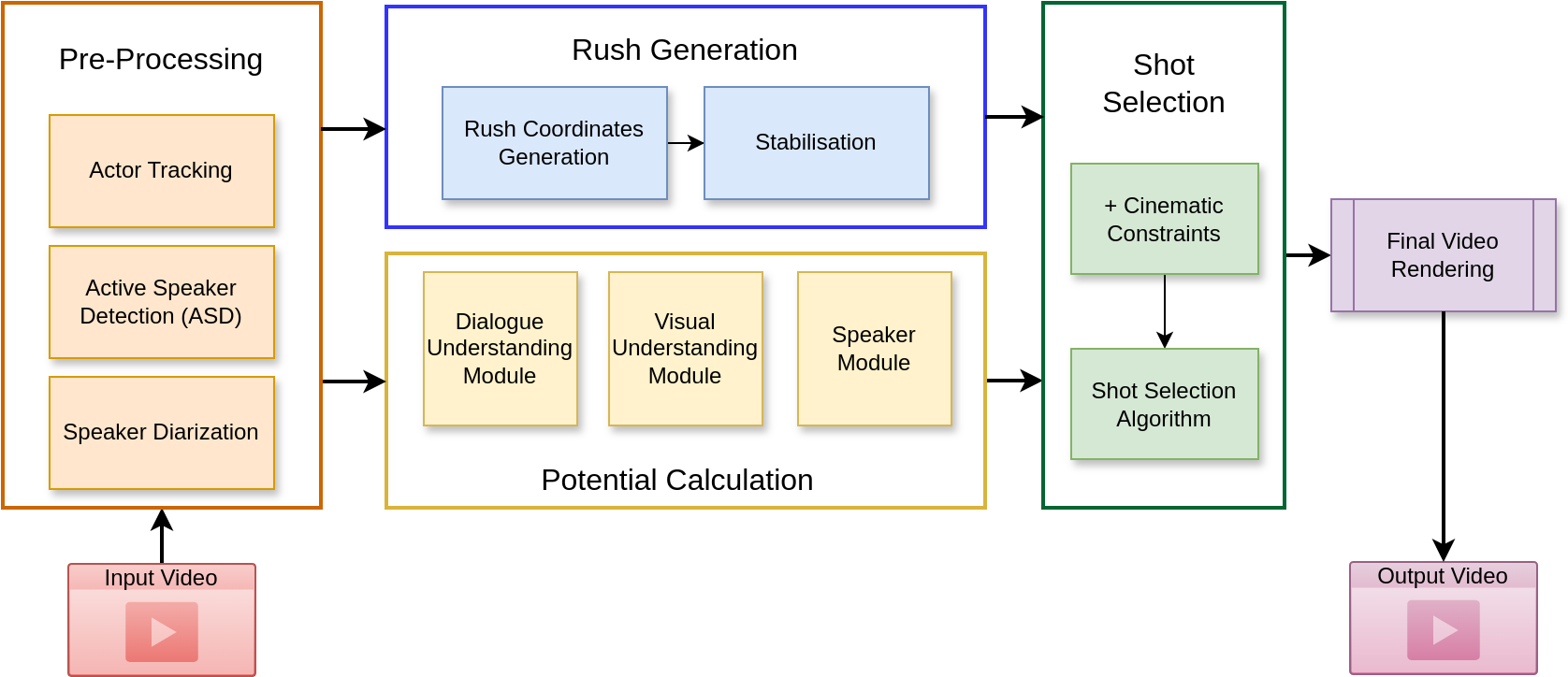}
\caption{\ourmethod~Pipeline: This fully automated pipeline takes input in the form of video and face crops + IDs and outputs the completely edited video. The various parts of the pipeline are shown in the figure, with each step operating on the outputs of the previous ones.}
\label{fig:pipeline}
\end{figure*}

\subsection{Automated Camera Selection}

Automated camera selection has been studied in 3D environments, particularly for applications in pre-visualization (previz) and computer games. Early research~\cite{christianson1996declarative, he1996virtual} utilized film idioms and conventional formulas for capturing scenes through sequences of shots~\cite{arijon1976grammar}. Another stream of research treats camera selection as a discrete optimization problem, addressing it through dynamic programming approaches~\cite{lino2011computational, galvane2015continuity, merabti2016virtual}. Meratbi~\etal~\cite{merabti2016virtual} limits to dialogue-driven scenes and utilizes Hidden Markov Models (HMM) for camera selection. Other important aspects, such as editing rhythm, the avoidance of jump cuts, and continuity editing, are addressed in~\cite{galvane2015continuity}. Although our work draws from these efforts, stage performances present unique limitations, including restricted camera placement and a lack of access to scene geometry, character localization, and event data available in 3D environments.

The problem of camera selection has been thoroughly studied in the context of sports events~\cite{wang2008automatic,chen2010personalized,ChenWHCSG13,chen2018camera,pan2021smart}. Wang~\etal~\cite{wang2008automatic} employs HMMs for the task, and salient action coverage is maximized in~\cite{chen2010personalized}. Other studies~\cite{ChenWHCSG13,chen2018camera} adopt a data-driven methodology, training regressors to evaluate the significance of each camera angle at any given moment. Pan~\etal~\cite{pan2021smart} employs an event-based approach, initially identifying events of interest before selecting the most appealing views for those events.

Arev ~\etal~\cite{arev2014automatic} propose a method for the automatic editing of multiple social camera feeds. Their approach uses a trellis graph representation to optimize an objective function, which seeks to maximize coverage of the key content in a scene while maintaining adherence to cinematographic principles, such as avoiding jump cuts. The importance of the content is quantified based on joint attention across multiple cameras~\cite{NIPS2012_1bf2efbb}. Work by Leake~\etal~\cite{leake-speaker} introduces an idiom-based method for editing dialogue-driven scenes. Their system accepts multiple camera angles, various takes, and the film script as inputs, generating the most informative set of shots for each line of dialogue. 

A key distinction of our work from the aforementioned methods is the lack of multiple manually operated camera feeds or takes; instead, we utilize a single wide-angle recording and virtually simulate cameras for editing. 
The most similar work to ours is GAZED~\cite{gazed}, which uses the human gaze to identify salient scene elements, assuming that actor tracks are available. In contrast, our approach eliminates reliance on gaze data and introduces a fully automated editing pipeline that does not require actor tracks or rule-based idioms. Human gaze provides a strong direct proxy for scene importance; in contrast, our method relies on trained machine learning models, tackling several predictive uncertainties faced while automation in real-world applications. 

\section{EditIQ Overview}
This paper introduces a comprehensive, end-to-end video editing system - \ourmethod~ - designed to automate the cinematic editing and video production process. The system is designed to process static, high-definition recordings of the scene intended for editing as its input and transform them into visually appealing edited videos adhering to the established cinematic principles. The overall architecture of the system mirrors traditional video production pipelines and is structured into four key stages: 
\aptLtoX{(i) \textit{Pre-processing} of the input video (ii) \textit{Rush Generation} to generate cinematically valid shots for different actors or elements in the scene. (iii) \textit{Potential Calculation} to calculate each shot importance based on factors like dialogue understanding and spatial saliency cues. (iv) \textit{Shot Selection} to choose the most appropriate shot at each moment to ensure engaging storytelling.}{
\begin{inlinelist}
    \item \textit{Pre-processing} of the input video
    \item \textit{Rush Generation} to generate cinematically valid shots for different actors or elements in the scene.
    \item \textit{Potential Calculation} to calculate each shot importance based on factors like dialogue understanding and spatial saliency cues.
    \item \textit{Shot Selection} to choose the most appropriate shot at each moment to ensure engaging storytelling.
\end{inlinelist}}

The various steps in the pipeline are shown in the \Cref{fig:pipeline}. We'll talk about each of the pipeline steps in detail in the following sub-sections. The inputs to the pipeline include: 
\aptLtoX{(i) High-definition video recording captured from stationary camera(s) covering the entire scene (ii) Brief scene description and cast information, including the names of the actors and a single photo of each.}{\begin{inlinelist}
    \item High-definition video recording captured from stationary camera(s) covering the entire scene
    \item Brief scene description and cast information, including the names of the actors and a single photo of each.
\end{inlinelist}}

\begin{figure*}[t]
\centering
\includegraphics[width=\textwidth]{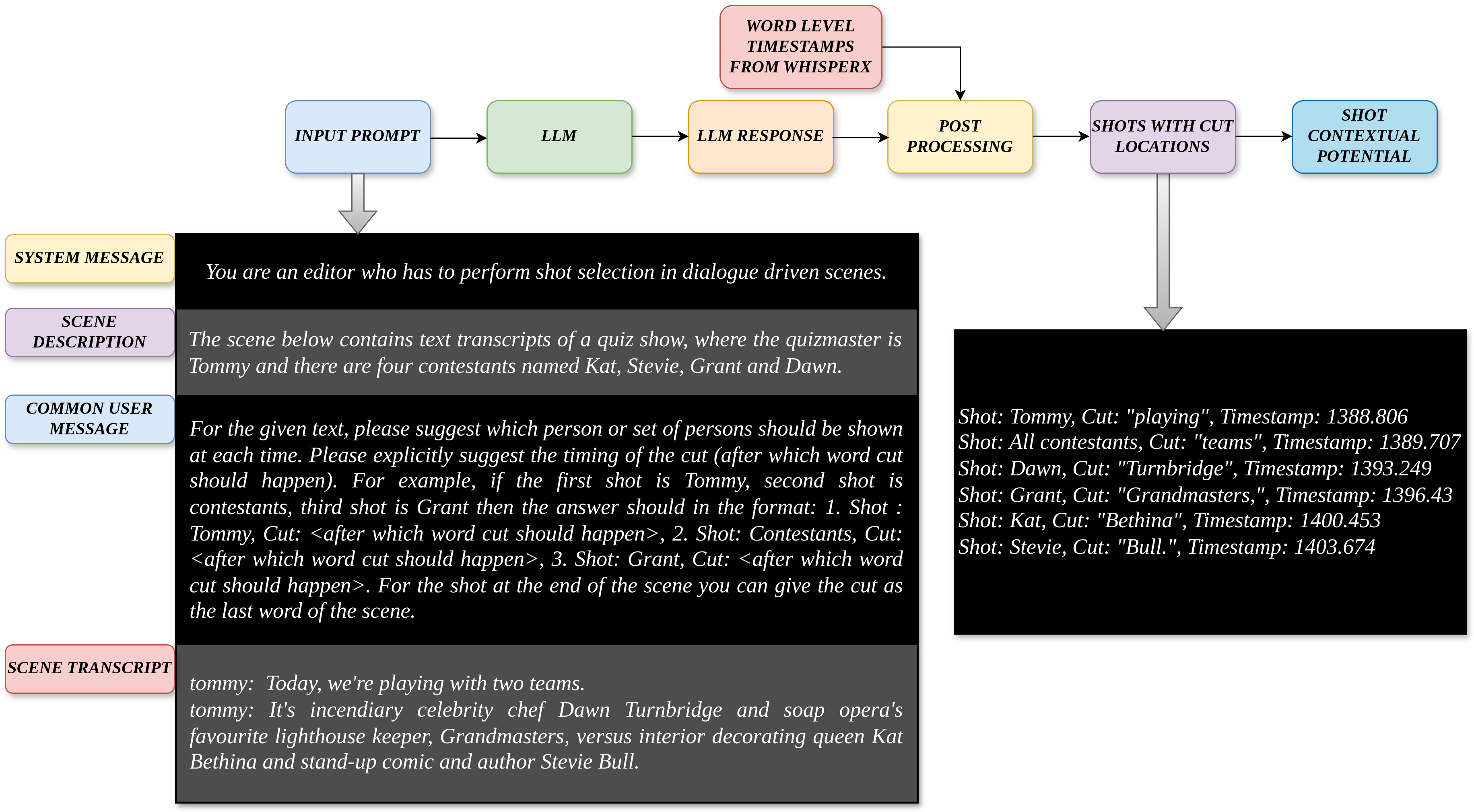}
\caption{Dialogue understanding module to get Contextual Potential from LLM for different shots based on the transcript of a scene. The post-processing in the above figure performs mapping between the LLM response and word level timestamps (from pre-processing) to get the cut locations.}
\label{fig:LLM_IO}
\end{figure*}

\subsection{Pre-Processing}
\label{sec:pre}
Hereafter, frames from the original wide-angle input video are also referred to as "\textit{master shots}". 
Given the master shot, several key features are derived to support the upcoming stages of the pipeline:

\begin{enumerate}
    \item \textit{Actor Tracks}: For the purposes of this work, we utilize the BoT-SORT \cite{botsort} model for person detection and tracking. This model provides a list of bounding box coordinates \begin{math}[x, y, h, w]\end{math} for each identified person in every frame, maintaining identity preserving tracks for each actor. Tracking errors, if any, were corrected before sending them forward. These tracks are essential for Shot Generation and Video Editing tasks. In practice, this step is algorithm-agnostic, enabling the use of any alternative algorithms.
    \item \textit{Character Aware Subtitling}:  We utilize the WhisperX~\cite{whisperx} model on the audio stream of our dataset to detect speech regions with word-level timestamps. The detected words are concatenated to generate a full transcription for each video, which is then segmented into sentences using a sentence tokenizer. Next, we employ the method from~\cite{korbar2023look} to generate character-aware subtitles, producing a complete dialogue transcript with accurate speech timestamps and speaker identification. In brief, we first pick high-quality exemplars for each character using TalkNet~\cite{talknet} and then leverage these exemplars to classify all speech segments by speaker identity \cite{Korbar22classification}.
    
\end{enumerate}

\subsection{Rush Generation}
The second stage of the \ourmethod~ pipeline consists of the shot generation task. Herein, a simulation approach~\cite{vcs} was employed to automatically produce virtual PTZ cameras by maneuvering multiple cropping windows following a particular actor or a group of actors within the master shot. We predominantly use medium shots (framing the actor from head to waist) for individual subjects and full shots (depicting the subject from head to toe) to capture two or more actors.

For a master shot with $n$ actors, we generate $2^n - 1$ virtual shots. This includes ${}^n\mathrm{C}_1$ 1-shots (individual actors), ${}^n\mathrm{C}_2$ 2-shots (two actors), ${}^n\mathrm{C}_3$ 3-shots, etc., all in a 16:9 aspect ratio. These shots, along with the master shot, are referred to as "rushes" for further selection and editing. We define $S_t$ as the set of rushes at time $t$ as:

\begin{equation}
\label{eqn:all_rushes}
\begin{split}
    S_t = \{ A \mid A \subseteq \{x_1, x_2, \dots, x_n\} \text{ and } A \neq \emptyset \} \cup \{\text{Master Shot}\} \\
    \text{ where }x_i \text{ is the } i^{th} \text{ actor}
\end{split}
\end{equation}

Smaller framings like Medium Shots (MS) highlight an actor's actions and expressions in detail, while larger framings like  Full Shots (FS) capture the whole actor(s), emphasizing their presence and context within the scene. An example of the generated rushes is presented in \Cref{fig:teaser}, featuring three actors. The rushes include the master shot, three 1-shots, two 2-shots, and one 3-shot.

In contrast to the previous works~\cite{vcs}, which use an upper-body detector, we use a person pose estimator, which gives us more control over the shot framing and provides a more detailed understanding of the subject’s orientation and body posture. 
For this study, we use YOLOv8-Pose \cite{yolov8, yolo_repo}, for its high precision and real-time performance.

Following~\cite{vcs}, we first obtain per frame shot estimations, which are then optimized to ensure well-composed shots mimicking the movement of professional cameramen. The virtual PTZ simulation is formulated as an optimization problem, with the objective of minimizing a sum-of-squares term that measures closeness to the original per frame estimations, combined with \( L_1 \)-norm regularization on velocity and jerk~\cite{vcs, 5995525}.

\begin{figure*}[t]
    \centering
    \includegraphics[width=0.8\linewidth]{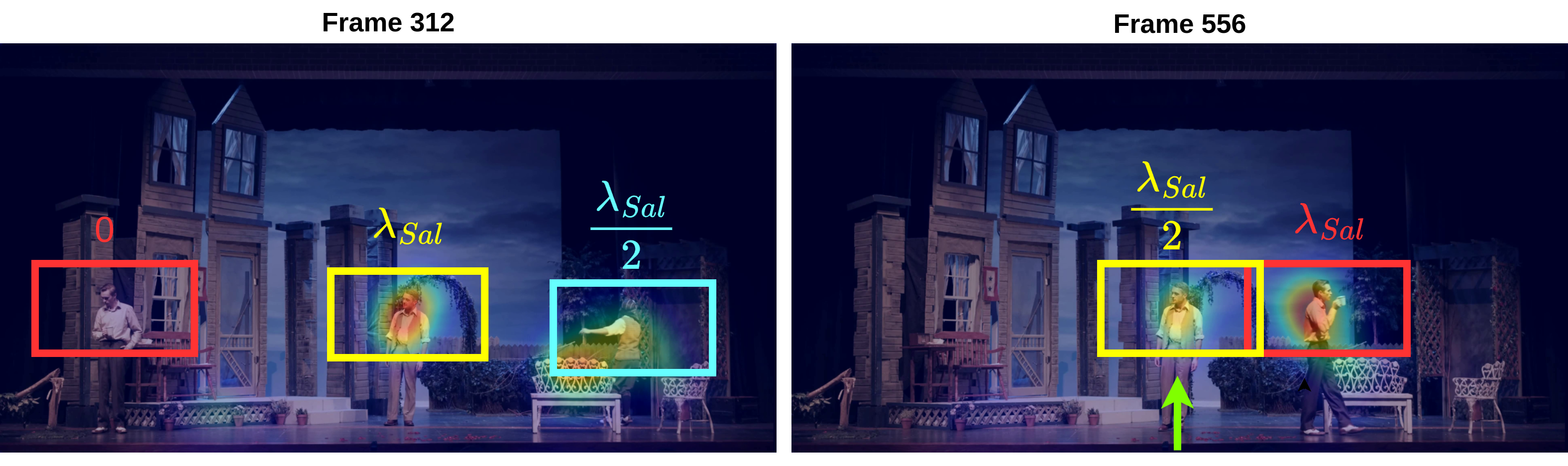}
    \caption{\textbf{Saliency potential} of different single-order shots for two frames in a theatre video (potential value is shown along with the actor shot). Green arrow indicates the speaker, if any.}
    \label{fig:saliency_potential}
\end{figure*}

\subsection{Dialogue Understanding Module — Contextual Potential}
\label{sec:LLM}

In video editing, dialogue is a core component that shapes how the story is told, characters are developed, and emotions are conveyed. An editor must not only understand the literal meaning of the words but also interpret the various attributes conveyed through dialogue, such as tone, emotion etc. By controlling when to cut, how to pace conversations, and which reactions to emphasize, editors can transform a simple dialogue scene into a powerful, emotionally engaging moment that drives the narrative forward and deepens the audience’s connection to the characters.

As part of the Dialogue Understanding Module, we utilize Large Language Models (LLMs) to interpret dialogues and assist with shot sequence suggestions. The LLM receives a scene transcript and a concise prompt with instructions for generating a shot sequence that visually narrates the scene. We also provide a brief scene description to enrich the context. The transcript includes speaker details (from the pre-processing step), allowing the LLM to identify the individual or group to be shown at each moment in the scene. Additionally, we request explicit cut points, specifying the exact word after which each cut should occur. The LLM response, containing shot suggestions and cut information, is post-processed with word-level timestamps to determine precise cut locations (as timestamps). Shot suggestions from LLMs are used to compute the contextual potential for each shot at any time instant.

~\Cref{fig:LLM_IO} illustrates this process using an example from the BBC-OSD, showing the user prompt, a sample transcript, and the final shot sequence suggestions with cut locations derived from word-level timestamps. A key principle followed for the contextual potential is that other shots that either include or overlap with the selected shot should not have a zero cost, but rather a small one, as they still partially capture the context used by the LLM during its recommendations.
The calculation of the contextual potential depends on the type of shot selected by the model:
\begin{enumerate}

    \item If a single-order shot is selected, it is assigned a cost of $\lambda_{c}$. Higher-order shots (p-order shots that contain the actor selected) are assigned a cost of $\frac{\lambda_{c}}{p}$, as they can also convey the same context as the single-order shot but less effectively than a close-up. All other shots receive a cost of $0$.

\begin{equation}
    C(s_{t}^{X}) =
    \begin{cases} 
        \lambda_{c} & \text{$s_{t}^{X}$ = $s_{t}^{X_{s}}$} \\   
        \frac{\lambda_{c}}{p} & \text{$x_{s}$ $\in$ X, \( |X| \) > 1}\\
        0 & \text{for remaining 1-shots}
    \end{cases}
 \label{eqn:ContextualPotential1}
\end{equation}

\noindent where $s_{t}^{X}$ refers to a shot $s$ at time $t$ that contains a set of actors $X$ and $s_{t}^{X_{s}}$ is the single-order shot selection from LLM with actor $x_{s}$.

    \item If a p-order shot ($p > 1$) is selected, 
    single-order shots involving actors from the selected shot are given a cost of $\frac{\lambda_{c}}{2^{p-1}}$. As calculated through ~\Cref{eqn:Higher-Order-Potential}, the selected shot receives a cost of $\lambda_{c}$, while all other shots receive a cost of less than $\lambda_{c}$.

\begin{equation}
    C(s_{t}^{X}) =
    \begin{cases} 
        \frac{\lambda_{c}}{2^{p-1}} &  \text{X $\in$ $X_{s}$, \( |X| \) = 1} \\ 
        0 & \text{for remaining 1-shots }
    \end{cases}
    \label{eqn:ContextualPotential2}
\end{equation}
where $s_{t}^{X}$ refers to a shot $s$ at time $t$ that contains a set of actors $X$ and $X_{s}$ is the higher-order shot selection from LLM with actors $X_{s} = \{ x_i \mid i = 1, 2, \ldots, p \}$

For higher-order shots
, the contextual potential is calculated as described in~\cite{gazed}. For example, consider a 2-shot $s_{t}^{X}$ where $X = \{x_{1},x_{2}\}$ containing actors $x_{1}$ and $x_{2}$. Contextual potential for this higher-order shot is defined in terms of the contextual potentials of the single-order shots $C(s_{t}^{x_{1}})$ and $C(s_{t}^{x_{2}})$ as follows:
\begin{equation}
    C(s_{t}^{\{x_{1},x_{2}\}}) = C(s_{t}^{x_{1}}) + C(s_{t}^{x_{2}}) - \left| C(s_{t}^{x_{1}}) - C(s_{t}^{x_{2}}) \right|
    \label{eqn:Higher-Order-Potential}
\end{equation}
%
Similarly, contextual potentials of two 2-shots \( C(s_{t}^{\{x_{1},x_{2}\}}) \) and \( C(s_{t}^{\{x_{2},x_{3}\}}) \) can be used to compute the contextual potential of a 3-shot \( C(s_{t}^{\{x_{1},x_{2},x_{3}\}}) \), when the actors appear on screen in the order \( x_{1} \), \( x_{2} \), \( x_{3} \) from left to right.
\end{enumerate}

\subsection{Visual Understanding Module — Saliency Potential}
\label{sec:Saliency}

As previously discussed, dialogue understanding is crucial in the video editing process. However, it faces a limitation in that contextual potential alone cannot fully capture the visual information present in a scene. To address this, we incorporate saliency prediction, which allows us to extract essential visual elements beyond the dialogue, emphasizing actions, reactions, and overall scene dynamics. This approach ensures a more comprehensive understanding of both verbal and non-verbal elements within the video.

The \ourmethod~ pipeline employs a modified 3D convolutional, visual-only \textit{Video Saliency Prediction} (VSP) model, building on the ViNet \cite{vinet} architecture. ViNet was chosen because of its compact size as compared to other VSP models yet having competitive results among other factors - code availability and reproducibility of results. We observe that the modified ViNet model more effectively captures the overall essence of the scene, beyond merely detecting motion cues or primary semantic features such as faces. Further details about the model and results are explained in the \Cref{appendix:saliency}.

The above VSP model outputs a saliency map for each frame of the video, equivalent to the size of the original video frame. We then apply a threshold of $\tau_{sal}$ on the map, to eliminate values below this threshold, enhancing the clarity of salient features while reducing noise.

Subsequently, the saliency score for each actor is computed by taking the mean of the thresholded saliency values within its bounding box. To ensure comparability between actors, we normalize these saliency scores across all actors. The actor with the highest saliency score is assigned a value of $\lambda_{Sal}$, while the second most salient actor receives a score of $\lambda_{Sal} / 2$:

\begin{equation}
    V(s_{t}^{x}) =
    \begin{cases} 
        \lambda_{Sal} & \text{$x$ is salient actor} \\
        \frac{\lambda_{Sal}}{2} & \text{$x$ is second-most salient actor} \\
        0 & \text{otherwise}
    \end{cases}
    \label{eqn:SaliencyPotenatial}
\end{equation}

where $s_{t}^{x}$ refers to a shot $s$ at time $t$ that contains a single actor $x$. An example of the same is illustrated in~\Cref{fig:saliency_potential}, we can observe that saliency potential picks actions and actor movements, rather than focusing solely on the speaker, and is especially effective when there are no dialogues. Saliency potential for higher-order shots is then computed from the saliency potentials of constituent lower-order shots similar to \Cref{eqn:Higher-Order-Potential}.

\subsection{Speaker Module — Speaker Potential}
\label{sec:Speaker}
Speaker potential is designed to assign importance to the shot corresponding to the active speaker. Given the speaker-aware subtitles and the corresponding timestamps, the speaker potential ($S$) is defined as follows:

    \begin{equation}
        S(s_{t}^{x}) =
        \begin{cases} 
            \lambda_{Sp} & \text{$x$ is speaker} \\
            0 & \text{otherwise}
        \end{cases}
        \label{eqn:SingleSpeakerPotenatial}
    \end{equation}

    where $s_{t}^{x}$ refers to a 1-shot containing actor $x$ at time $t$.

\subsection{Cinematic Constraints}
\label{sec:constraints}

While the aforementioned potentials provide costs based on the importance of various shots, editing decisions made solely on these costs may lack cinematic validity. The contextual potential derived from LLMs does not incorporate visual information, which can lead to issues like overlapping cuts or jump cuts. Additionally, LLMs lack awareness of the duration of the shot, as this aspect is addressed later during the post-processing phase in the Dialogue Understanding Module (\Cref{sec:LLM}). Similarly, saliency in its raw form is not designed for a delicate task like video editing and does not account for cinematic rules such as minimum shot duration or the continuity of flow. Therefore, it is crucial for any video editing pipeline to consider cinematic principles separately alongside these potentials or costs.

In this study, we adopt an approach similar to GAZED~\cite{gazed}, where penalties are applied to shots that violate various cinematic principles. These penalties are then fed into the Shot Selection Algorithm, which selects the final shots based on both the shot potentials and the associated penalties. We define four types of penalty terms, which are detailed below. The total penalty for any given shot is the cumulative sum of these individual penalties at any given timestamp.

\subsubsection{Overlap Penalty:}

During the transition between two consecutive shots, excessive overlap can lead to a jump cut, which disrupts the continuity and can be visually jarring. To avoid such visually jarring transitions we use overlap penalty, designed to minimize overlap between consecutive shots. This penalty is applied only when two distinct shots, \(s_{t}^{i}\) and \(s_{t+1}^{j}\), are involved and a cut occurs between them.

\[
O(s_{t}^{i}, s_{t+1}^{j}, \gamma) =
\begin{cases} 
0 & \text{if } \gamma \leq \alpha \\
\frac{\mu\gamma}{\alpha} & \text{if } \alpha < \gamma \leq \beta \\
\nu & \text{if } \gamma > \beta
\end{cases}
\]

Here, \(\gamma\) represents the overlap ratio, calculated as the intersection-over-union (IoU) between two consecutive shots \(s_{t}^{i}\) and \(s_{t+1}^{j}\). The penalty is piecewise, i.e. no penalty is applied if the IoU is below threshold \(\alpha\), a linear penalty is applied for IoU values between \(\alpha\) and \(\beta\), and a high penalty \(\nu\) is applied when the overlap ratio exceeds \(\beta\), showcasing a significant overlap that violates cinematic principles.

\subsubsection{Misframing Penalty:}

Poorly framed shots occur when another actor is partially visible in the current frame, which can disrupt the composition. For example, the shot suggestions provided by LLMs may not account for the spatial arrangement of actors, potentially recommending a cut to an actor sitting in close proximity to another. To avoid such shots, we define misframing penalty as follows:

\[
M(s_{t}^{i}) =
\begin{cases}
\lambda_{\text{mis}} & \text{if the framing is poor} \\
0 & \text{otherwise}
\end{cases}
\]

If a shot \(s_{t}^{i}\) is found to be poorly framed, the penalty \(\lambda_{\text{mis}}\) is added to its cost. A framing is defined to be poor if it overlaps with actors beyond the shot definition.

\subsubsection{Rhythm Penalty:}
The pacing of cuts plays an important part in determining the overall feel of a scene in video editing. Shot duration directly influences how some audiences perceive the mood and energy of a sequence. For example, longer shots create a slower rhythm, bringing in calmness or emotional depth, often used in romantic or contemplative scenes. While, on the other hand, shorter shots create a faster rhythm, heightening tension or energy, a technique commonly applied in action scenes in editing. To manage the rhythm of cuts, we use the rhythm penalty, which regulates shot duration to maintain cinematic flow. The rhythm penalty is applied based on the duration of the current shot ( $\tau$ ), calculated as follows:

\[
R(s_{t}^{i}, s_{t-1}^{j}, \tau) =
\begin{cases} 
\gamma_1 \left( 1 - \frac{1}{1 + \exp (l - \tau)} \right) & \text{if } i \neq j \\
\gamma_2 \left( 1 - \frac{1}{1 + \exp (- m + \tau)} \right) & \text{if } i = j
\end{cases}
\]

In this equation, \(\tau\) is the time the current shot has been held, and \(l\) and \(m\) are parameters that control the rhythm timings. The constants \(\gamma_1\) and \(\gamma_2\) are scaling factors for rhythm penalty. When transitioning to a new shot \( (i \neq j) \), the penalty increases if the new shot is cut too quickly, i.e., before \(\tau = l\) seconds, to prevent rapid cutting. While on the other hand, if a shot is held for too long \( (i = j) \), a penalty builds up as \(\tau\) exceeds \(m\) seconds, encouraging a cut to introduce new visual information. Together, these two conditions help control the rhythm of cuts, ensuring that the scene is neither too rushed nor overly static.

\subsubsection{Transition Penalty:}

Extremely fast cuts in editing can confuse or disorient the audience and can undermine the emotional weight of a scene. Fast cuts can obscure important details and weaken storytelling clarity, and might compromise aesthetic quality. To prevent this, we apply a transition penalty to promote a minimum shot duration. Given two consecutive shots, \(s_{t}^{i}\) and \(s_{t+1}^{j}\), the penalty is defined as:

\[
T(s_{t}^{i}, s_{t+1}^{j}) =
\begin{cases}
0 & \text{if } i = j \\
\lambda_{trans} & \text{if } i \neq j
\end{cases}
\]

Here, \(\lambda_{trans}\) is the transition penalty parameter.

\subsection{Shot Selection}\label{shot_selection} 
Given the multiple types of shots and rushes, our next step in the \ourmethod~ pipeline involves selecting the shot that best fits the storytelling at each moment in time. 
We frame shot selection as a discrete optimization problem, evaluating the importance of each shot per frame while adhering to cinematic principles like avoiding jump cuts, rapid transitions, and irregular cutting rhythms. Shot importance at each time is determined by the potentials as explained in \Cref{sec:LLM}, \Cref{sec:Saliency}, and \Cref{sec:Speaker}, while cinematic principles are incorporated as penalty terms. The final solution is derived by finding the optimal path in an editing graph, which, for a scene with \(n\) actors, consists of \(2^n - 1\) nodes per frame. Each node represents a rush, with edges indicating transitions (cuts) or continuity (no cut) between shots.

Given a sequence of frames $t = [1,2, \dots, T]$ and the set of generated shots (rushes) $S_{t}$ (\Cref{eqn:all_rushes}), our method selects a sequence of shots $\epsilon = \{ r_{t} \mid i = 1, 2, \ldots, T \}$, $r_{t} \in S_{t}$ by minimizing the following objective function:

\begin{equation}
\label{eqn:objectivefunction}
\begin{split}
    E(\epsilon) = \sum_{t=1}^{T} -\ln\Big(U(r_{t})\Big) + \sum_{t=2}^{T} \Big[O(r_{t-1}, r_{t}, \gamma) + R(r_{t}, r_{t-1}, \tau) + \\ T(r_{t-1}, r_{t})\Big]  + \sum_{t=1}^{T} M(r_{t})
\end{split}
\end{equation}

where $U(r_{t})$ is the unary cost for a shot, representing the shot's importance. This unary cost is the cumulative sum of contextual potential, saliency potential, and speaker potential (\Cref{eqn:unarypotential}). The second and third terms represent different penalties described in \Cref{sec:constraints}.

\begin{equation}
\label{eqn:unarypotential}
    U(r_{t}) = C(r_{t}) + V(r_{t}) + S(r_{t})
\end{equation}

We solve ~\Cref{eqn:objectivefunction} using dynamic programming. Our method outputs a sequence of shots for each frame $t$ selected from a series of shots generated over time $\{ S_{t} \mid i = 1, 2, \ldots, T \}$. We build a cost matrix $CM(r_{t},t)$, $r_{t} \in S_{t}, t = [1,2, \dots, T]$ whose elements are computed recursively as follows:

\begin{displaymath}
\begin{split}
       CM(r_{t},t)  = 
    \begin{cases}
        -\ln(U(r_{t})) + M(r_{t}) & t = 1 \\
         \min_{k} \Big[ CM(r_{k}, t-1) - \ln(U(r_{t})) \\ + O(r_{k}, r_{t}, \gamma) + R(r_{t}, r_{k}, \tau) \\+ T(r_{k}, r_{t}) + M(r_{t}) \Big] & \text{otherwise}
    \end{cases}
\end{split}
\end{displaymath}


The cost matrix is constructed during a forward pass across the time dimension. For each element in the matrix, we calculate and store the minimum cost required to reach it. Once the matrix is completed, we backtrack to determine a sequence of optimal shots. For the edited video, we use the wide shot or master shot from the original footage as the establishing shot, setting its duration to 2 seconds and then optimizing only over the remaining frames.

\section{Experiments}

To assess whether our method - \ourmethod, which leverages a dialogue understanding module and saliency prediction to guide shot selection, results in a visually compelling and coherent cinematic representation of scenes, we conducted a psychophysical user study involving twenty participants, with the details as follows.

\subsection{Dataset}

The study utilizes the BBC Old School Dataset (BBC-OSD) \cite{bbc_osd}.
BBC-OSD, curated by BBC R\&D, is a comprehensive resource for advancing research into AI-driven automated video editing. The dataset includes comedy fiction (sitcom), drama, and game show elements and is set during the filming of a fictional game show called "Old School". It includes raw footage of multiple takes from the short TV program, along with behind-the-scenes content and rich metadata. Unlike conventional TV shoots, this production was tailored specifically to create data for automated editing, offering static wide-angle views and multi-participant interactions. The dataset provides insights into the entire production process, from planning to metadata generation, enabling the development of sophisticated AI editing systems for various use cases.
They also provide a human-edited programme as a benchmark for automated editing systems. We use the Edit Decision List (EDL) corresponding to the human-edited programme to extract videos from the raw footage, with a total duration of approximately 30 minutes.

In addition to the above-defined dataset, we selected eleven segments from three stage and theatre performances recorded in 4K resolution ($3840 \times 2160$). These videos include a mix of music concerts and various theatre acts, all captured using a wide-angle static camera, with no pan, cut, or zoom operations. The purpose of selecting these recordings is to evaluate the effectiveness of our \ourmethod~editing pipeline on more diverse scenarios. The chosen videos present a range of challenging cases, including rapid dialogues, actor co-referencing, abrupt story transitions, and critical background actions and emotions. Each video requires precise editing to ensure the narrative flows smoothly without missing key elements or important shots.

\subsection{LLM Configuration \& Details}

For our LLM-based inferences, we utilized the Claude 3.5 Sonnet model developed by Anthropic, specifically the "claude-3-5-sonnet-20240620" checkpoint ~\cite{Anthropic_2024}. The maximum context window length for the Claude model was 200K tokens during the time of building the system. For reproducibility purposes, we set the temperature parameter to 0.

\subsection{Parameter Selection}

Cinematic constraint parameters play a crucial role in shaping the output of the pipeline and can be seen as the personalization of the edits. Most parameters in \ourmethod~ are either drawn from established literature or set empirically. For instance, the rhythm penalty parameter $m$, which governs the maximum sequence length, is set to $7$, reflecting the average shot length in films over the past two decades \cite{shot-duration}. The minimum shot duration, controlled by parameter $l$, is set to $1$, with \(\gamma_1\) assigned a high value (100), as cuts shorter than 1 second tend to disrupt continuity, and very fast cuts are generally undesirable.  Similarly, the penalty for overlap, $\nu$ is kept at a very high value ($10^{6}$), to strictly avoid jump cuts. For overlap cost parameters, we set $\alpha = 0.15$, as cuts with less than 15\% overlap typically pose no visual issues, while $\beta = 0.3$ ensures that cuts with over 30\% overlap are flagged as abrupt. 

These parameters can be adjusted to customize the editing style, such as creating faster or slower-paced edits. The algorithm's computational efficiency enables interactive content exploration, allowing for real-time adjustments to personalize the final output.

\subsection{Baselines}

We compared the videos generated using the \ourmethod~ pipeline against various competing video editing baselines, including Random, Wide, and Speaker. These baselines were selected because they do not require any manual data collection for the editing process. For a fair comparison, all baseline videos were shown to users at the same resolution and audio quality, with each video retaining the exact timeline of the original footage.

In addition to these baselines, we included and conducted ablation studies using only the LLM-based Contextual Potential and Saliency-based Visual Potential to demonstrate that relying on these methods alone is insufficient for effective video editing. Furthermore, we included Human Edits as a baseline for the BBC-OSD Dataset, allowing us to compare our results with professional editing standards.

\subsubsection{Random Baseline:}

The Random baseline (Ran) is a simple method where shots are chosen randomly from the available footage at different time intervals, without considering what is happening in the scene (the context). After selecting the shots, we apply cinematic rules and penalties to try and improve the video, since random selections often break these rules and can lead to awkward or unappealing results. This lack of coherence makes it the weakest baseline in terms of narrative flow and visual consistency.

\subsubsection{Wide Baseline:}

The Wide baseline is inspired by video re-targeting techniques and is similar to the letterboxing method described in prior research \cite{gaze-video-retargeting}. This approach focuses on selecting the widest possible shot that includes all performers on stage. This shot is essentially a zoomed-in version of the master shot, capturing the entire scene without excluding any actors. The goal is to ensure that no one is left out of the frame, prioritizing coverage over more focused or dynamic shots. This method is simple and effective for keeping all performers in view at all times, but it lacks the flexibility to adapt to changes in action or focus within the scene.

\subsubsection{Speaker Baseline:}

Speaker cues are valuable for editing dialogue driven scenes, as highlighted in previous studies by Ranjan \etal~\cite{ranjan-speaker} and Leake \etal~\cite{leake-speaker}, who advocate for selecting shots that clearly showcase the speaker. Our speaker-based (Sp) editing baseline follows a similar approach by choosing the shot that best highlights the speaker from the available footage. This selection process relies on information obtained from the character-aware subtitling (\Cref{sec:pre}). The current shot selection remains unchanged until a different speaker takes the floor. To minimize abrupt transitions, a minimum shot duration is enforced. If there is a period of silence lasting more than 10 seconds, the algorithm will switch to a wide shot for the subsequent time interval.

\subsubsection{LLM-Only baseline:}

The LLM-based baseline leverages insights from large language models to select shots that align with the narrative context. By analyzing the dialogues within the video, this approach aims to choose shots that enhance storytelling and maintain coherence. The LLM Potential is computed as outlined in \Cref{sec:LLM}, ensuring that the selected shots contribute meaningfully to the narrative. However, since LLMs lack information about long or short cut times (based on how contextual potential is calculated) as well as lack of visual information (overlapping shots, jump cuts), corrections based on cinematic principles are applied afterward to ensure a smooth visual flow and adherence to established editing standards. This process helps mitigate potential issues that may arise from the initial shot selection, enhancing the overall quality of the edited video.

\subsubsection{Saliency-Only baseline:}

The Saliency (Sal) baseline uses visual saliency detection to find the most important parts of a video frame, focusing on shots that highlight these key elements. By emphasizing what draws the viewer's attention, this method aims to create a more engaging experience. The Saliency Potential is calculated as outlined in \Cref{sec:Saliency}, which measures how well each shot showcases these important visuals. However, since saliency detection doesn’t consider rules like minimum shot duration or storytelling flow, we apply corrections based on cinematic principles afterwards.

\subsubsection{Human Edits (Only for BBC-OSD):}

These edits were performed by professional video editors at the BBC, providing a valuable point of comparison for our approach against an actual edit (or an established ground truth). Users rated these edits alongside others, unaware that they were created by humans. The videos used for this baseline were taken directly from the BBC-OSD without any modifications, ensuring an accurate representation of professional editing standards.

\begin{figure*}[htbp]
    \centering
    \includegraphics[width=\linewidth]{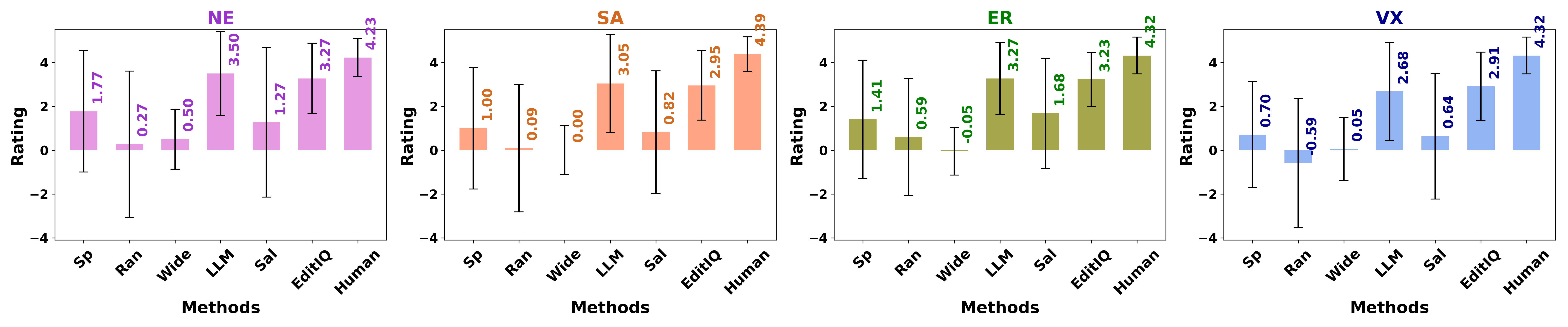}
    \includegraphics[width=\linewidth]{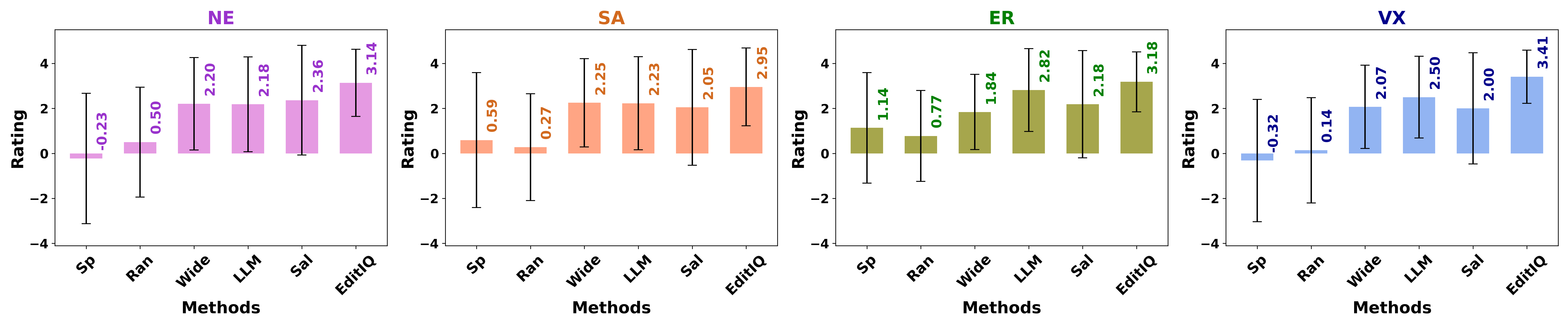}
    \caption{\textbf{User Study Evaluation:} Bar plots denoting mean user ratings for the different editing methodologies across four evaluation attributes for the (top row) BBC-OSD and (bottom row) Theatre recordings. Error bars denote unit standard deviation. Best viewed in color and under zoom.}
    \label{fig:US_plots}
\end{figure*}

\section{Evaluation \& User Study}
\begin{sloppypar}
\subsection{Materials \& Methods:} To evaluate the efficacy of \ourmethod~ against the aforementioned video editing baselines, we conducted a psychophysical study involving 20 users (aged 20--25 and including 2 females). Original and edited versions of 11 videos from the BBC-OSD and 11 theatre performance videos generated by all baselines plus \ourmethod~ were viewed by users. The maximum video length over these 22 videos was 94s. For a fair comparison, identical \ourmethod~ parameters were used for generating all video edits. Upon viewing the original video, each user viewed the edited\footnote{Corresponding to the six baselines including human-edited plus \ourmethod~ for BBC-OSD, and five baselines plus \ourmethod~ for theatre videos.} versions at the same pixel resolution in a random sequence to eliminate order-specific effects. 

The study design was such that each user viewed the original and edited versions of 2-3 videos so that the 20 users cumulatively viewed all 22 recordings, and the experiment lasted around 20 minutes per user. We ensured that the original and edited versions of each recording received exactly two user ratings resulting in a $11~(\text{video types}) \times \text{2~(user ratings/video)} \times 7~\text{(editing strategies)}$ factor design for BBC-OSD, and a $11~(\text{video types}) \times \text{2~(user ratings/video)} \times 6~\text{(editing methods)}$ factor design for theatre recording videos. 
\end{sloppypar}

{Users were naive to the strategies employed for generating the edited versions. Users had to \textit{compare} each edited version against the original and provide a Likert rating on a [-5,5] scale for each of the attributes described below. These attributes were adopted from~\cite{rachavarapu2018watch,gaze-video-retargeting}, and are designed to evaluate how effectively the edited versions capture \textit{focal scene events} given event recording constraints. The attributes of interest included:}
 
\begin{itemize}
 \item[(1)] \textbf{Narrational Effectiveness (NE):} \textit{How effectively did the edited video convey the original narrative?}
 \item[(2)] \textbf{Scene actions (SA):} \textit{How well did the edited video capture actor movements and actions?}
 \item[(3)] \textbf{Actor Emotions and Reactions (ER):} \textit{How well did the edited video capture actor emotions and reactions?}
 \item[(4)] \textbf{Viewing experience (VX):} \textit{How would you rate the edited video for aesthetic quality?}
\end{itemize}

Users were familiarized regarding these attributes, and about cinematic video editing conventions prior to the study. Users had to rate for questions (1)--(4), \textit{relative} to a reference score of `0' for the original video. A \textit{positive} score would therefore imply that the edited version was \textit{better} than the original for the target attribute, while a \textit{negative} score conveyed that the edited version was \textit{worse} than the original with respect to the criterion. User responses were collated, and mean scores were computed per criterion and editing strategy over all videos (see~\Cref{fig:US_plots}). Statistics and inferences from the user study are presented below.

\subsection{Results and Discussion}
\subsubsection{BBC Old School (BBC-OSD):} Bar plots depicting mean user scores across attributes and editing methods are presented in \Cref{fig:US_plots}. A two-way balanced analysis of variance (ANOVA) on the compiled NE, SA, ER, and VX user scores across methods revealed the main effects of \textit{editing strategy} on user opinions ($p < 0.000001$ for all four attributes), and \textit{video type} ($p < 0.005$ for all four attributes). No interaction effects were noted. We hypothesized that combining LLM and visual saliency cues via \ourmethod~ would result in an engaging, vivid, and aesthetic edit, which is generally validated by~\Cref{fig:US_plots}. Across the four attributes, \ourmethod~ expectedly outperforms the Random, Speaker, Wide, and Saliency baselines, but performs comparably to LLM and inferior to professional human edits.

{Investigating specific attributes, \textit{post-hoc} independent $t$-tests on NE scores revealed a significant difference between \ourmethod~ vs. Sp ($p < 0.05$), \ourmethod~ vs. Ran ($p < 0.001$), \ourmethod~ vs. Wide ($p < 0.000001$) and \ourmethod~ vs. Sal ($p < 0.05$). \ourmethod~ vs. LLM NE values were very comparable ($p = 0.6728$), while NE scores for human edits were significantly higher than for \ourmethod~ ($p < 0.05$). These results cumulatively convey that carefully compositing shots which provide a closer view of the key scene actor(s) and action(s) is crucial for effective scene narration. The Random baseline, which selects shots independent of scene content, performs worst, followed by the wide baseline, which can only present the entire scene context without a focus on scene details. The Speaker and Saliency-based editing methods, which respectively employ speech and visual scene cues for shot selections, perform comparably ($p = 0.5968$), but LLM-based editing, which is guided by the scene narrative, significantly outperforms visual saliency ($p<0.05$).} conveying that visual cues only supplement LLM capabilities for automated video editing. 

For conveying scene actions, user score trends are very similar to NE scores. \ourmethod~significantly outperforms Ran ($p<0.0005$), Wide ($p<0.000001$), Sp ($p<0.001$) and Sal ($p<0.05$), while performing similar to LLM ($p=0.9177$) and under-performing compared to human editing ($p<0.005$). Visual saliency and speech-based editing are deemed comparable by users ($p=0.7306$), while LLM-based editing outperforms both methods ($p<0.05$). Slightly different trends are, however, noted with respect to conveying actor emotions and reactions. \ourmethod~is rated significantly higher than Ran ($p<0.0005$), Wide ($p<0.000001$), Sp ($p<0.001$) and Sal ($p<0.05$), very comparable to LLM ($p=0.9177$) and inferior to human editing ($p<0.005$). Saliency-based editing scores better for this attribute, performing better than Sp ($p<0.01$), but still lower than LLM-based editing ($p<0.05$). These trends convey that saliency is more effective at capturing visually prominent facial expression changes and reactions, even if they cannot effectively capture the general scene narrative or critical scene actions. 

Finally, familiar trends repeat with respect to the viewing experience. \ourmethod~outperforms all other competing automated approaches, but scores significantly lower than professional human editing ($p<0.001$). Saliency and speaker-based editing again perform very comparably ($p= 0.9326$), with LLM-based editing outperforming these two approaches ($p<0.05$).

\subsubsection{Theatre recordings:} Scores very different to the BBC-OSD, which captures a quiz event involving four participants in a smallish venue, are obtained for the theatre recordings capturing a larger venue, and where actions and events can possibly happen at the stage periphery as well. Human-edited outputs are not available for the theatre sequences, and therefore we will only compare the automated editing approaches.

Repeating a two-way ANOVA for theatre sequence user scores conveyed the main effect of editing methodology ($p<0.0005$ for all attributes), but no effect of video type or interaction effects. Given the large spatial context in theatre performances, the Wide baseline capturing the entire scene scores relatively higher as compared to the BBC-OSD.    

For narrational effectiveness, the Sp and Ran baselines score similarly poorly ($p=0.3731$), while the Wide, LLM and Sal editing approaches perform superiorly and comparably. \ourmethod~ achieves the highest scores, which are significantly higher than for Sp ($p<0.0001$), Ran ($p<0.0005$), marginally higher than Wide ($p=0.0927$) and LLM ($p=0.0909$) and insignificantly higher than the Sal baseline. With respect to scene actions, trends are generally similar, with \ourmethod~ scoring significantly higher than Sp ($p<0.005$) and Ran ($p<0.0005$) but only insignificantly higher than the Wide, LLM, and Sal baselines. For facial expressions and reactions, \ourmethod~ again scores significantly higher than Ran ($p<0.00005$), Sp ($p<0.005$) and Wide ($p<0.01$), marginally higher than Sal ($p=0.095$) but only insignificantly higher than LLM-based editing. Finally, with respect to viewing experience, our \ourmethod~ approach scores significantly higher than Sp ($p<0.00001$), Ran ($p<0.000005$), Wide ($p<0.001$), and Sal ($p<0.05$) baselines, and marginally higher than the LLM ($p=0.0571$) baseline. This conveys that our proposed editing approach can effectively combine LLM and visual-based cues to engagingly and vividly convey static camera recordings to viewers.

\subsubsection{Past-Experience of Participants} 
Among the 20 participants, five had prior experience in video editing, while the remaining individuals lacked familiarity with video editing tools and techniques. Experienced participants exhibited a more critical perspective, consistently assigning significantly lower ratings to the random baseline compared to non-experienced participants, whose scores were closer to neutral. This highlights their heightened ability to detect and penalize editing flaws. Additionally, experienced participants demonstrated greater appreciation for high-quality edits, rating human edits higher—4.4 (NE), 4.8 (SA), 4.6 (ER), 5.0 (VX)—compared to non-experienced participants—4.17 (NE), 4.26 (SA), 4.23 (ER), 4.11 (VX). This indicates that their expertise allowed them to recognize the nuances and limitations inherent in the editing process, leading to a more informed evaluation.




\subsubsection{Discussion Summary} We evaluated \ourmethod~ against competing baselines under two varied settings: (1) a quiz event captured by the BBC-OSD, and (2) theatre recordings that involve a very diverse context and dynamics compared to quizzes. Although the combination of the saliency and LLM cues is not very beneficial in the quiz context, where speech cues essentially guide visual attention, the benefit is more apparent for theatre performance edits where actions from actors other than the speaker could be regarded as salient. In both settings, however, \ourmethod~ is found to generally outperform other automated approaches while only scoring inferior with respect to professional human editing for BBC-OSD.

\section{Conclusion}

This work introduces \ourmethod, a framework for the automatic editing of stage performance videos captured by unmanned, static, wide-angle, high-resolution cameras. We employ LLMs for dialogue understanding and prompt it to suggest which person or set of persons should be shown at each word timestamp of the character-aware subtitles. Additionally, we employ video saliency prediction methods to capture actions and other visual elements that are not conveyed through the dialogue. The LLM suggestions, saliency predictions, and speaker information are combined together to quantify the importance of each shot in the generated rushes at each time. These unary shot potentials are then combined with cinematic penalties like avoiding jump cuts and fast cuts, avoiding improper framings, and maintaining rhythm. The result is a meticulously edited sequence that not only preserves key content but also adheres to cinematic principles, resulting in a visually compelling video. The effectiveness of \ourmethod~compared to competing baselines is demonstrated through a psychophysical study involving twenty participants using the BBC-OSD and eleven theatre performance videos. \ourmethod~generally outperforms other baselines, scoring lower only in comparison to professional human editing for the BBC-OSD.

\section{Limitations and Ethical Considerations}

A key limitation of the current system is its inability to perform real-time editing, a critical feature for live events. Previous work has established the feasibility of online rush generation, stabilization, and camera selection~\cite{achary2024real, vcs}. Future endeavors will seek to integrate these with a streaming LLM variant, enabling dialogues to be processed incrementally rather than as a complete script.

%
%

It is also important to emphasize that this project is not intended to replace human editors but rather to serve as an assistive tool. The results from the human evaluation clearly demonstrate that human edits consistently received higher scores than those produced by automated methods, underscoring the irreplaceable expertise of professional editors. Instead, this system aims to support editors by generating novel ideas or reducing their workload, particularly in tasks such as selecting shots from extensive footage. Nevertheless, the system provides a cost-effective solution for low-budget theaters, allowing them to create visually appealing edits of performances without the need for expensive multi-camera setups or professional editors.

\begin{acks}

We express our sincere gratitude to BBC R\&D and the authors of the Old School Dataset (OSD), which includes raw rushes and expertly crafted human edits. This meticulously curated dataset has been pivotal in shaping our work. We further extend special thanks to Stephen Jolly and Graeme Phillipson for their dedicated efforts in formally providing us with this resource. We would also like to thank The Dorset Players for their performance of \textit{The 39 Steps} theatre play and to Martin HS Theatre for their presentation of \textit{All My Sons} play. We also thank all participants who contributed to the User Study.


\end{acks}

\bibliographystyle{ACM-Reference-Format}
\bibliography{main-citations}

\clearpage
\appendix

\begin{figure*}[hbt!]
\centering
\includegraphics[width=\textwidth]{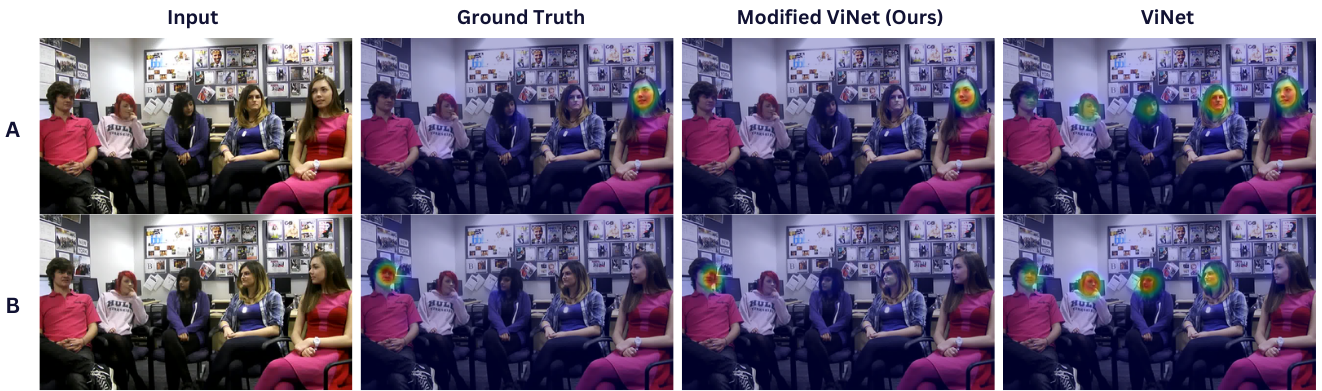}
\caption{Here, we compare our modified Saliency Prediction model with state-of-the-art ViNet Model~\cite{vinet}. Our model captures the essence of the whole scene and performs joint attention to capture interactions. It focuses on the key actor, whereas ViNet limits to head movements and captures all the faces as salient.}
\label{fig:saliency}
\end{figure*}

\section{Prompts}
\label{appendix:prompts}

\subsection{System Message}
\texttt{You are an editor who has to perform shot selection in dialogue driven scenes.}

\subsection{Scene Description}
\subsubsection{BBC-OSD}
\texttt{The scene below contains text transcripts of a quiz show, where the quizmaster is Tommy and there are four contestants named Kat, Stevie, Grant and Dawn.}

\subsubsection{Theatre}
\texttt{The scene below contains text transcripts of a scene from a theatre play.}

\subsection{Common Prompt}
\subsubsection{BBC-OSD}
\texttt{For the given text, please suggest which person or set of persons should be shown at each time. Please explicitly suggest the timing of the cut (after which word cut should happen). For example, if the first shot is Tommy, second shot is contestants, third shot is Grant then the answer should in the format: 1. Shot: Tommy, Cut: <after which word cut should happen>, 2. Shot: Contestants, Cut: <after which word cut should happen>, 3. Shot: Grant, Cut: <after which word cut should happen>. For the shot at the end of the scene you can give the cut as the last word of the scene.}

\subsubsection{Theatre}
\texttt{For the given text, please suggest which person or set of actors should be shown at each time. Please explicitly suggest the timing of the cut (after which word cut should happen). For example, if the first shot is actorX, second shot is (actorX and actorY), third shot is actorZ then the answer should in the format: 1. Shot: actorX, Cut: <after which word cut should happen>, 2. Shot: (actorX and actorY), Cut: <after which word cut should happen>, 3. Shot: actorZ, Cut: <after which word cut should happen>. For the shot at the end of the scene you can give the cut as the last word of the scene.}

\section{Video Saliency Prediction (VSP) Model}
\label{appendix:saliency}

We use a modified VSP model based on the 3D convolutional ViNet~\cite{vinet} model, with two key modifications made to the original ViNet~\cite{vinet} model: \begin{enumerate}
    \item We addressed limitations found in action classification backbones like S3D~\cite{s3d}, which tend to overlook background actions by focusing on primary motion in human-centric videos. Instead, we integrated a Spatio-Temporal Action Localization (STAL) backbone~\cite{acarnet, slowfast}, pre-trained on the AVA actions dataset~\cite{gu2018ava}, alongside our custom decoder. This combination improves the ability to localize and classify actions, thereby better capturing the essence of the scene.
    \item The ViNet decoder was restructured to enhance computational efficiency, by incorporating filter groups~\cite{8100116} and channel shuffle layers~\cite{zhang2018shufflenet}. This method reduces the original model’s size and parameter count by threefold, while simultaneously improving Saliency Prediction performance.
\end{enumerate}

We observe that the modified ViNet model captures the overall essence of the scene more effectively, beyond merely detecting motion cues or primary semantic features such as faces. For instance, \Cref{fig:saliency} illustrates a frame from a video in the MVVA~\cite{mvva} dataset, where a group of people are being interviewed. While other existing SOTA models, like the original ViNet~\cite{vinet}, limit to head movements and end up mistakenly highlighting all faces as salient, our model accurately identifies the most relevant face, such as the person speaking or the one receiving attention from others in the scene. Results for the model on a few human-centric datasets are shown in ~\Cref{tab:mvva_results} \& ~\Cref{tab:coutrot_etmd}.

\begin{table*}[h]
    \centering
    \caption{Results on MVVA~\cite{mvva} Dataset}
    \begin{tabular}{|c||c c c c|}
    \hline
    \bfseries METHOD & \multicolumn{4}{c|}{\bfseries MVVA}\\
       & CC$\uparrow$ & NSS$\uparrow$ & AUC$\uparrow$ & KLDiv$\downarrow$ \\
    \hline\hline
    \bfseries TASED-Net~\cite{min2019tased} & 0.653 & 3.319 & 0.905 & 0.970\\
    \bfseries STAViS~\cite{tsiami2020stavis} & 0.77 & 3.060 & 0.91 & 0.80\\   
    \bfseries ViNet~\cite{vinet} & 0.81 & 4.470 & \textcolor{red}{\underline{0.93}} & 0.75\\
    \bfseries VAM-Net~\cite{vamnet} & 0.741 & 4.002 & 0.912 & 0.783\\
    \hline
    \bfseries Ours & \textcolor{red}{\underline{0.821}} & \textcolor{red}{\underline{4.792}} & \textcolor{red}{\underline{0.93}} & \textcolor{red}{\underline{0.689}}\\      
    \hline
    \end{tabular}
\label{tab:mvva_results}
\end{table*}

\begin{table*}[h]
\centering
\caption{Results on Coutrot2~\cite{coutrot2} and ETMD~\cite{etmd} Datasets}
\label{tab:coutrot_etmd}
\begin{tabular}{|c||c c c c||c c c c|}
    \hline
    \bfseries METHOD & \multicolumn{4}{|c||}{\bfseries Coutrot2} & \multicolumn{4}{c|}{\bfseries ETMD} \\
       & CC$\uparrow$    & NSS$\uparrow$   & AUC$\uparrow$ & SIM$\uparrow$   & CC$\uparrow$    & NSS$\uparrow$  & AUC$\uparrow$ & SIM$\uparrow$ \\
    \hline\hline
\bfseries TASED-Net~\cite{min2019tased} & 0.437 & 3.17  & 0.921 & 0.314 & 0.509 & 2.63 & 0.916 & 0.366 \\        
\bfseries STAViS~\cite{tsiami2020stavis}    & 0.652 & 4.19  & 0.940 & 0.447 & 0.560 & 2.84 & 0.929 & 0.412 \\         
\bfseries ViNet~\cite{vinet}     & 0.724 & 5.61  & 0.95  & 0.466 & 0.569 & 3.06 & 0.928 & 0.409 \\
\bfseries TSFP-Net~\cite{tsfpnet}  & 0.718 & 5.30  & 0.957 & 0.516 & 0.576 & 3.09 & 0.932 & 0.433 \\
\bfseries CASP-Net~\cite{xiong2023casp}  & 0.756 & 6.07  & 0.963 & 0.567 & 0.616 & 3.31 & 0.938 & 0.471 \\               
\hline 
\bfseries Ours      & \textcolor{red}{\underline{0.860}} & \textcolor{red}{\underline{6.563}} & \textcolor{red}{\underline{0.963}} & \textcolor{red}{\underline{0.610}} & \textcolor{red}{\underline{0.632}} & \textcolor{red}{\underline{3.519}} & \textcolor{red}{\underline{0.943}} & \textcolor{red}{\underline{0.493}} \\
\hline
\end{tabular}
\end{table*}

\end{document}